# Missing author address information in Web of Science
## 一An explorative study

Weishu Liu[1], Guangyuan Hu[2], Li Tang*

1. School of Information Management and Engineering, Zhejiang University of Finance and Economics, Hangzhou, China

2. School of Public Economics and Administration, Shanghai University of Finance and Economics, Shanghai, China

3. School of International Relations and Public Affairs, Fudan University, Shanghai, China  Email: Litang@fudan.edu.cn

**Abstract:** Bibliometric analysis is increasingly used to evaluate and compare research performance across geographical regions. However, the problem of missing information from author addresses has not attracted sufficient attention from scholars and practitioners. This study probes the missing data problem in the three core journal citation databases of Web of Science (WoS). Our findings reveal that from 1900 to 2015 over one-fifth of the publications indexed in WoS have completely missing information from the address field. The magnitude of the problem varies greatly among time periods, citation databases, document types, and publishing languages. The problem is especially serious for research in the sciences and social sciences published before the early 1970s and remains significant for recent publications in the arts and humanities. Further examinations suggest that many records with completely missing address information do not represent scholarly research. Full-text scanning of a random sample reveals that about 40% of the articles have some address information that is not indexed in WoS. This study also finds that the problem of partially missing address information for U.S. research has diminished dramatically since 1998. The paper ends by providing some discussion and tentative remedies.

**Keywords:** information omission; author address; research evaluation

1. Introduction

Credible data and appropriate model specification are the very two pillars of a solid empirical study (Heckman, 2005; Young & Holsteen, 2017). This also holds true for informetrics research (Waltman & Eck, 2012; Wang & Shapira, 2011; Tang & Walsh, 2010). The author's address is a critical element for analyzing research performance and competitiveness, as spatially relevant information such as the author's affiliation, region, and country/territory information all come from the address field provided in the bylines of publications (Csomós, 2017; Sun & Grimes, 2016; Yu, 2015; Zhou, Thijs, & Glanzel, 2009). In a similar vein, studies on research collaboration across organizations, regions, and countries also rely heavily on authors' addresses (Guan, Yan, & Zhang, 2015; Lemarchand, 2012; Li & Li, 2015; Perianes-Rodriguez, Waltman, & van Eck, 2016; Wang, Wang, & Philipsen, 2017). Although a high presence rate of author address information is the foundation of various bibliometric analyses, the problem of missing data is often neglected. Take *Journal of Informetrics* (JOI) and *Journal of the Association for Information Science and Technology*

*(JASIST)*, two top journals in library and information science, for example. Over the last two years (2015–2016), at least 20 articles utilize the address fields of Web of Science, one of the most widely used publication databases for research evaluation,[1] but only two articles mention the information omission problem. Neither discusses the potential impacts of such omission on their findings, nor do they address how to remedy this problem.

Such omission-induced errors have raised concerns of some bibliometricians. For instance, Marx (2011) points out that there are substantial data missing from the field of author address in Web of Science (hereinafter WoS) for papers published in 1973. Jacsó (2009) finds that about 14% of records indexed in WoS have no author country information for the publication years of 1980 to 2009. Even for publications in 2006–2015, research finds that 5% of the records in Science Citation Index Expanded (hereinafter SCIE), 9% of the records in Social Sciences Citation Index (hereinafter SSCI), and 42% of the records in Arts & Humanities Citation Index (hereinafter A&HCI) have no author address information reported (Liu, Ding, & Gu, 2017).

This high absence rate of authors' address information casts shadows on the validity of address-derived indicators in bibliometric analysis. Yet little attention is paid to the status quo and dynamics of this missing data, as well as the sources of this problem. To fill in some knowledge in this research gap, this study explores the real situation of the problem of missing data from the author address field in WoS. The structure of the rest of this paper is as follows. In Section 2 we delineate our data and methods. In the analysis section, we depict the problem of completely missing data from the address field. This work pays special attention to the characteristics of missing information across different time periods, citation databases, document types, publishing languages, and journals. We next construct a random sample of publications with completely missing data from the address field and investigate the sources of missing information. This research also investigates partially missing address information, a lesser problem of data omission, using U.S. publications as a special illustrative example. The paper ends with a conclusion, limitations, and tentative suggestions for remedying this problem.

2. Data and methods

We chose the three core journal citation indexes of WoS—SCIE, SSCI, and A&HCI—to explore the problem of missing address information. It is to the credit of the Century of Science$^{TM}$ and the Century of Social Sciences$^{TM}$ Initiatives (Thomson Reuters, 2009) that studies are able to examine the evolution of scientific advancement from a historical perspective with time spans extended from 1900 to 2015 for SCIE and SSCI and from 1975 to 2015 for A&HCI (Chen and Ho, 2015; Liu, Zhang, and Hong, 2011; Liu et al., 2012).

We adopted the following three queries to retrieve records without author address information. Query #1 returns hits of all records indexed in the WoS database, and Query #2 retrieves records reporting author address information. The difference of the two retrieved subsets, i.e., Query # 3, is the target of this study: the records with

---

[1] When considering all document types, there are 188 and 446 papers published in JOI and JASIST, respectively, from 2015 to 2016. During this period, at least 8 JOI papers and 13 JASIST papers utilized the address fields of WoS, and only two mentioned this problem. Data assessed on Jan. 23, 2017.

missing information from the author address field.[2]

Query #1: PY=(1900-2015)

Query #2: AD=(A* OR B* OR C* OR D* OR E* OR F* OR G* OR H* OR I* OR J* OR K* OR L* OR M* OR N* OR O* OR P* OR Q* OR R* OR S* OR T* OR U* OR V* OR W* OR X* OR Y* OR Z* OR 0* OR 1* OR 2* OR 3* OR 4* OR 5* OR 6* OR 7* OR 8* OR 9*)

Query #3: (Query #1) NOT (Query #2)

Figure 1 is a flowchart of our data retrievals and analyses on information omission in the address field. As illustrated, we next probe the characteristics of completely missing address information by time dynamics, document type, publishing language and journal distribution. Words in grey boxes explain the selected time frames for analysis. Based on a random sample we then examine the sources of completely missing data. We also investigate the time dynamics of the problem of partially missing address information in the case of U.S. publications.

(Insert Figure 1 here)

Figure 1 Flowchart of data analysis on missing address information

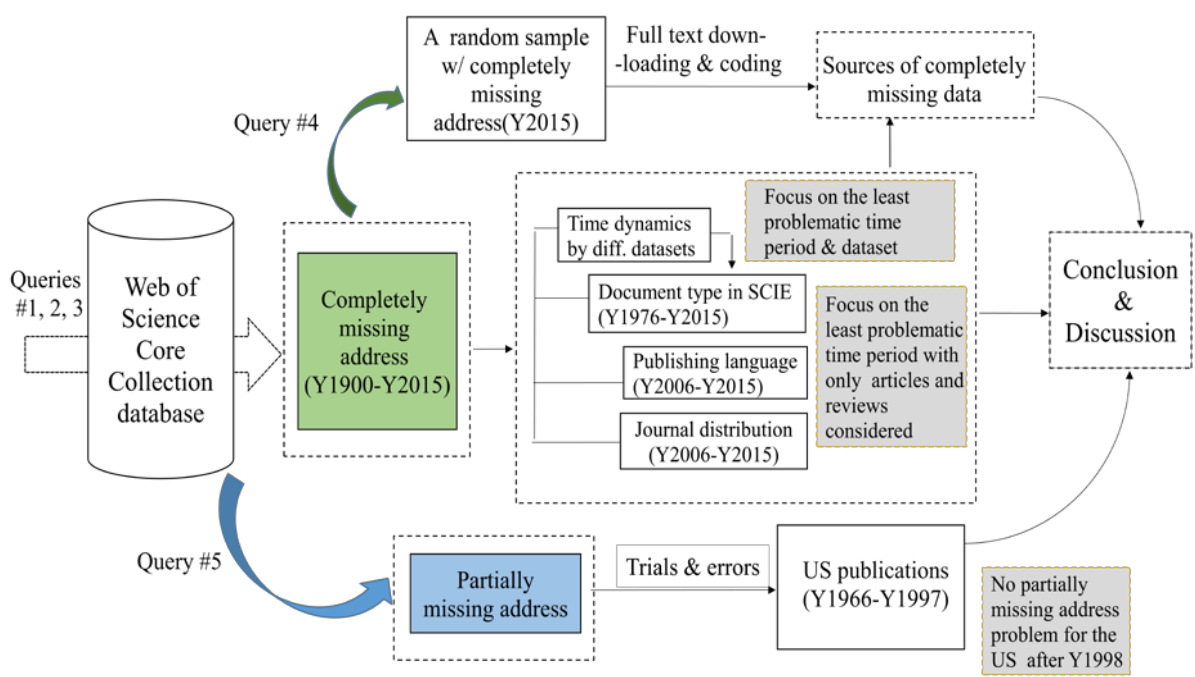

3. Analyses

There are two types of missing address information. One is completely missing data from the whole field of author address, and the other is partial information, such as country name, missing from the author address field. In this paper we focus on the former case, the more serious problem, while also discussing partially missing information at the end of the analysis section.

---

[2] We accessed these data through the Library of Xi'an Jiao Tong University on 15 November 2016.

## 3.1 Quantity and quality

During our examined 116 years (Y1900-Y2015), the SCIE, SSCI, and A&HCI databases indexed over 55 million publications. More than one-fifth of the records have no information provided in the author address field. Among them, about 41% classify as the document type "article." That is to say, 5.14 million original articles indexed in WoS have data missing from the address field.

In addition to the sheer number of records with missing data, further examination suggests that the quality of these records is not negligible, either. A large number of these articles without address information come from prestigious journals such as *Nature* (47,225 articles), *Lancet* (46,503 articles), and *Science* (24,232 articles). Among the 199 publications cited over 10,000 times in the WoS Core Collection, 52 publications, i.e., over one-quarter of the most influential research, have no author address available in the databases.

Some may argue that cumulative citations favor older articles. That is true: 54 out of 199 articles with 10,000+ citations were published prior to 1973, and 170 were published before 2000. But the problem of missing information does persist among highly cited articles in more recent years. Our data show that 28 publications published within the years 2006–2015 and cited more than 1,000 times have no address information. These influential publications can be invisible in bibliometric studies if researchers implement improper searching or data cleaning.

## 3.2 Time dynamics

As noted previously, over the period of 1900 to 2015, more than 12 million publications indexed in WoS have no author address information. An interesting question arises: is the missing address rate steady over time, or can it be ignored in recent years? We probe this question in SCIE, SSCI, and A&HCI database separately.

*SCIE database*

SCIE indexed more than 46 million records published from 1900 to 2015. Among them, approximately 8.26 million (about 18%) report no author address. In order to depict the changing pattern of these records, Panel A of Figure 2 shows both annual number and relative share of records without author address information.

(Insert Figure 2 here)

Thanks to the Century of Science Initiative (Thomson Reuters, 2009), nearly one million (969,861) records published from 1900 to 1944 are now back filed in the SCIE. However, 726,725 of them (about 75%) include no author address information. During this period, the annual number of missing address records, ranging between 9,000 and 25,000, is relatively stable. The missing address rate started at 81% in 1900, peaked at 91% in 1904, and kept a decreasing trend since then to about 60% in the early 1940s.

The period from 1945 to 1972 is the most worrisome. Among 4.5 million SCIE indexed papers, about 4.4 million records (or about 98%) lack author address information. That is to say, only around 2% of SCIE records during this period have author address information in the database. Luckily, the SCIE database began to index author address information systematically in 1973. The average missing author address rate dropped to about 7.6% from 1973 to 2015, which is much lower than the previous two periods. But still, among the 40.5 million articles published in 1973–2015, there are about 3.1 million scientific publications indexed in SCIE reporting no address information.

*SSCI database*

The story is similar for the SSCI database. From 1900 to 2015, about 2.45 million out of 8 million publications (i.e., 30%) have no author address information. As demonstrated in Panel B of Figure 2, the changing pattern of these records naturally divides into four successive phases: 1900–1955, 1956–1965, 1966–1972, and 1973–2015.

Figure 2 Time dynamics of records with address information missing from WoS Core Collection database: 1900–2015

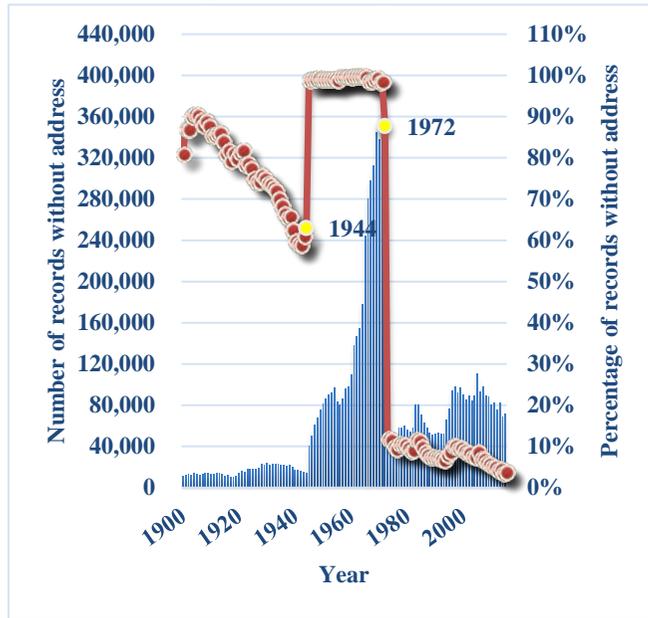
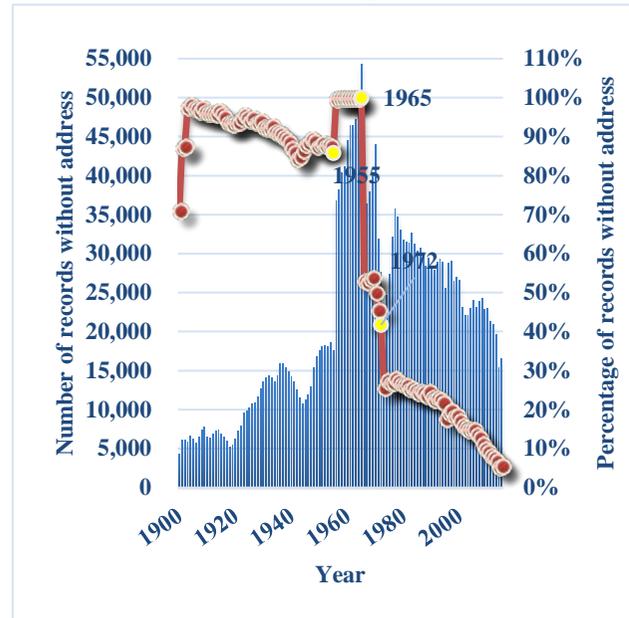
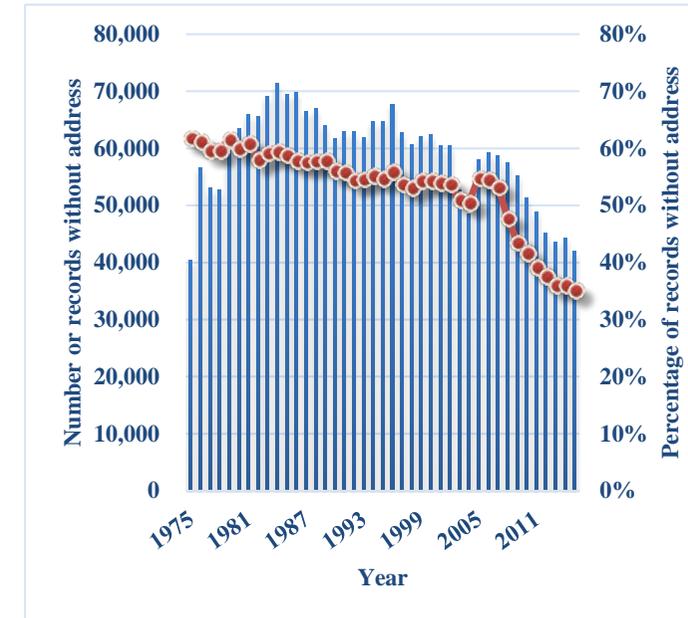

Panel A: SCIE                               Panel B: SSCI                               Panel C: A&HCI

Notes: Data accessed on 15 November 2016.
All document types are included. The blue bars refer to the numbers of publications with data missing from the address field, while the red dots refer to missing address rates, which equal the number of missing address records divided by the total number of records per year.

Over the period of 1900–1955, 669 thousand records were indexed in SSCI, but with 607 thousand records (about 90%) having no information in the address field. The most troublesome period is from 1956 to 1965: almost none of the records (44 out of 445,638 SSCI indexed publications) contain author address information. The situation has improved since 1966. The missing address rate was 42% in 1972. It then plummeted to 26% in 1973 and finally to 6% in 2015.

*A&HCI database*

The A&HCI database started indexing publications in arts and humanities in 1975. Our data show that from 1975 to 2015 A&HCI indexed over 4.5 million records. About 2.4 million records are missing author address information. As depicted in Panel C of Figure 2, different from the SCIE and SSCI databases, the annual number of records without address information is relatively stable during the whole period, with a slight decreasing trend over the last decade. About 62% of the total records in 1975 have no address information, and the missing data proportion is still as high as 35% in 2015.

3.3 Document type

The previous section shows that the missing author address rates are high before 1976 for all three citation indexes. The missing address rates have decreased in recent years but remain problematic in SSCI and are especially troublesome in the A&HCI database. Zooming in on the 40-year period of 1976–2015, we next examine the distribution of records with missing address data by document type.

There are roughly 40 different types of documents indexed in the three core citation indexes of WoS. But in reality, bibliometric analysis often limits the document type to substantive document types such as original articles and reviews (Bornmann & Bauer, 2015; Tang, Shapira, & Youtie, 2015; Waltman, Tijssen, & Eck, 2011; Waltman & Eck, 2012). As illustrated in Figure 3, over the period of 1976–2015 when all document types are considered (blue dots), the missing data rates, ranging from 4% to 65%, are consistently higher than those of original articles (red dots) and reviews (gray dots). In other words, the shares of original papers and reviews with missing address information are not as alarming as it may seem; this is particularly true for the SCIE in most recent years.

(Insert Figure 3 here)

Then what types of documents drive the high missing address data?

Focusing on SCIE, the least problematic database of the address-missing issue among the WoS Core Collection database, Table 1 lays out the top ten document types in terms of the number of records with missing address information over two time periods: 1976–2015 and 2006–2015.

As shown, original article tops the list with the largest number of records among different document types: 2.4% of original articles (670 thousand papers) published in 1976–2015 have no address information in SCIE. In the latest ten-year period of 2006–2015, though, 89 thousand articles and reviews have no address information in SCIE. Both smaller proportions and descending ranks of missing data suggest the problem has diminished. While this is good news, bibliometric analyses often include papers published in several decades. Therefore, papers published in earlier decades

Figure 3 Numbers and rates of data missing from the address field by document type: 1976–2015

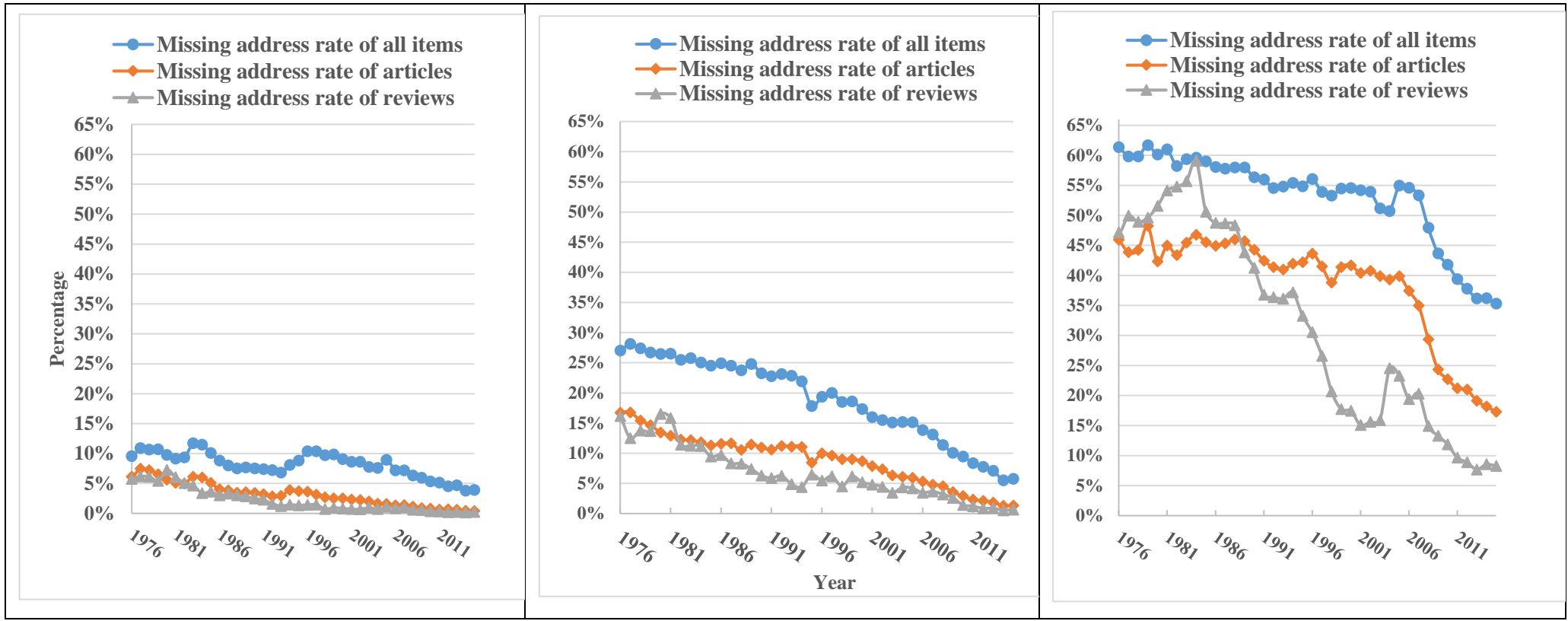

Panel A: SCIE  Panel B: SSCI  Panel C: A&HCI

Notes: Data accessed on 15 November 2016.
Blue dots refer to missing address rates for all document types. Red dots refer to missing address rates for original articles. Gray dots are missing address rates for reviews.
Missing address rate of a specific document type = Number of missing address records of this specific document type / Total number of records of this specific document type.

with information missing from the address field may still be cause for concern when considering the analyses' results.

A common practice in bibliometric-based research evaluations is to exclude other document types such as news items, book reviews, and editorial material in the analyzing sample. From the perspective of reducing omission-induced errors, this is also a good thing. On the other hand, considering the share and volume of original articles and reviews in WoS, Figure 3 and Table 1 suggest that the problem of missing address information should not be ignored when considering only original articles for research evaluation. This is particularly true for articles indexed in A&HCI in recent years.

(Insert Table 1 here)

Table 1 Numbers and rates of data missing from the address field by document type: The case of SCIE

| Period | 1976–2015 | | | 2006–2015 | | |
|---|---|---|---|---|---|---|
| Rank | Document type | Records without address | Missing address rate | Document type | Records without address | Missing address rate |
| 1 | **Article** | **670,569** | **2.4%** | Meeting abstract | 246,018 | 9.8% |
| 2 | Editorial material | 670,001 | 39.4% | Editorial material | 199,222 | 26.2% |
| 3 | Meeting abstract | 535,851 | 9.3% | News item | 175,734 | 90.6% |
| 4 | News item | 403,629 | 93.8% | **Article** | **86,513** | **0.8%** |
| 5 | Letter | 267,459 | 18.8% | Correction | 72,790 | 61.2% |
| 6 | Correction | 210,089 | 74.8% | Letter | 49,784 | 12.4% |
| 7 | Note | 71,703 | 8.1% | Biographical item | 21,323 | 57.0% |
| 8 | Book review | 50,601 | 34.0% | Book review | 6,650 | 20.0% |
| 9 | Biographical item | 43,526 | 60.8% | Proceedings paper | 2,710 | 0.5% |
| 10 | Item about an individual | 24587 | 68.5% | **Review** | **2,660** | **0.4%** |

Notes: Data accessed on 2 April 2018.

Missing address rate of a specific document type = Number of missing address records of this document type in SCIE / Total number of records of this document type in SCIE within the same time period.

3.4 Publishing language

Publishing language has been factored into recent bibliometric studies on research impacts (Egghe, Rousseau, & Yitzhaki, 1999; Reguant, 1994; Liu, 2017), yet it remains unknown which language has the largest amount of information missing from the address field. As shown in Table 2, from 2006 to 2015, there are 89,000 publications reporting no address information.[3] Among them, research written in English, the lingua franca of scholarly communication, has the largest number of articles and reviews with missing address information in WoS. Of all the publications without address information, 85% are in English, 6% in German, and 3% in French. Chinese is the third largest publishing language in the SCIE database (69,038 articles), and only 0.2% (136) of articles written in Chinese have no address information reported. If we normalize the ratio by the indexed publications in individual languages, Dutch seems to stand out: 49.6% of SCIE papers written in Dutch have the problem of missing information in the address field. Taking a closer look into these Dutch papers, we find that 638 out of 643 (over 99%) are published in one journal, *Tijdschrift Voor Diergeneeskunde*, a trade magazine. This magazine's nonscholarly nature, to some extent, relieves our concern of evaluation inaccuracy induced by missing address information related to Dutch publications.[4]

---
[3] We only consider original articles and reviews here.

SSCI and A&HCI tell a similar but more problematic story. From 2006 to 2015, approximately 2.7% of SSCI publications and 23.3% of A&HCI articles have no address information. English is the publishing language with the largest number of publications with the problem of missing address information in both data sources, but Russian papers (about 1/3 missing address rate in SSCI and 2/3 missing address rate in A&HCI), German papers (9.1% in SSCI and 51.7% in A&HCI), and French papers (10.6% in SSCI and 46.7% in A&HCI) are among the most worrisome. These figures warn us that when both language and address are examined in a study, information omission cannot be ignored. Previous studies show the language bias of WoS favoring English articles (Lin and Zhang, 2007). This may be not a bad thing considering the sheer size of records with missing data in other languages such as French and German.

(Insert Table 2 here)

3.5 Journal distribution

Table 3 identifies the top journals with the largest numbers of records with missing address information during the period of 2006–2015, which comprised 18%, 24%, and 14% of publications with completely missing addresses indexed in the SCI, SSCI, and A&HCI databases, respectively.

(Insert Table 3 here)

As shown in Table 3, the majority of these journals are low-impact journals. Seventeen out of twenty journals belong to the bottom 25% quartile of the impact factor distribution in 2016 Journal Citation Reports.[5] Table 3 also indicates that except for the French journal *Positif* with 100% missing address information, the remaining 29 journals have different proportions of publications with address information indexed in WoS ranging from 27.8% to 99.9%. This finding suggests that there exist inconsistent journal policies on address inclusion. In alignment with the findings shown in Table 2, non-English journals suffer from high missing address rates, while some English journals are also not free of address omission.

People may suspect nonscientific journals or magazines to lead in missing address information.[6] Utilizing Ulrich's Knowledgebase, we are able to identify the scientific

---

[4] There is no agreed upon judgment on the "scholarly nature" of WoS indexed publications. The criteria we use is to see whether a publication satisfies all of the following three conditions: 1) the document type of the publication is "article" or "review", 2) the journal information of the serial type retrieved from Ulrich's Knowledgebase is "journal" rather than "magazine", and 3) the content type retrieved from Ulrich's Knowledgebase is "academic" rather than "consumer" or "trade".

[5] No information is available on impact factors for journals indexed in A&HCI. For journals no longer indexed in WoS, we chose their latest possible journal impact factors and associated quartile values.

[6] We would like to thank an anonymous reviewer for directing us to think about the content type of our analyzing sample.

Table 2 Missing address rates by publishing language and citation index: 2006–2015

| | SCIE | | | | SSCI | | | | A&HCI | | | |
|---|---|---|---|---|---|---|---|---|---|---|---|---|
| Rank | Language | Missing address | Total records | % | Language | Missing address | Total records | % | Language | Missing address | Total records | % |
| 1 | English | 75,822 | 11,360,801 | 0.7 | English | 35,471 | 1,528,527 | 2.3 | English | 52,120 | 315,392 | 16.5 |
| 2 | German | 4,964 | 70,859 | 7.0 | **Russian** | **1,979** | **5,987** | **33.1** | **French** | **14,998** | **32,104** | **46.7** |
| 3 | French | 2,298 | 54,349 | 4.2 | German | 1,743 | 19,126 | 9.1 | **German** | **12,224** | **23,625** | **51.7** |
| 4 | Spanish | 1,204 | 45,598 | 2.6 | Spanish | 1,109 | 22,827 | 4.9 | **Italian** | **9,715** | **13,679** | **71.0** |
| 5 | Japanese | 955 | 14,243 | 6.7 | French | 995 | 9,372 | 10.6 | **Russian** | **4,526** | **6,693** | **67.6** |
| 6 | Russian | 689 | 12,755 | 5.4 | Portuguese | 309 | 10,880 | 2.8 | Spanish | 3,633 | 20,771 | 17.5 |
| 7 | **Dutch** | **643** | **1,296** | **49.6** | **Norwegian** | **256** | **489** | **52.4** | **Czech** | **450** | **2,066** | **21.8** |
| 8 | Portuguese | 534 | 44,367 | 1.2 | **Hungarian** | **197** | **433** | **45.5** | **Croatian** | **408** | **1,521** | **26.8** |
| 9 | Polish | 380 | 15,747 | 2.4 | Turkish | 143 | 2,569 | 5.6 | Portuguese | 375 | 2,993 | 12.5 |
| 10 | Croatian | 316 | 1,828 | 17.3 | Croatian | 122 | 991 | 12.3 | **Dutch** | **338** | **1,538** | **22.0** |
| | Total | 89,003 | 11,719,195 | 0.8 | Total | 42,917 | 1,609,961 | 2.7 | Total | 100,252 | 430,541 | 23.3 |

Notes: Data accessed on 15 November 2016.

Only articles and reviews considered.

Table 3 Ranking of journals by the number of records with missing address information: 2006–2015

| | Journal | Serial type [a] | Content type [b] | Language [c] | #Missing address | #Total | Missing address rate % | JIF quartile |
|---|---|---|---|---|---|---|---|---|
| SCI | Chemical & Engineering News | Magazine | Trade | English | 3438 | 4017 | 85.6 | Q4 |
| | Oil & Gas Journal | Magazine | Trade | English | 2689 | 3802 | 70.7 | Q4 |
| | New Scientist | Magazine | Consumer | English | 2202 | 2428 | 90.7 | Q4 |
| | Naval Architect | Journal | Academic | English | 1721 | 1759 | 97.8 | Q4 |
| | Genetic Engineering & Biotechnology News | Magazine | Trade | English | 1430 | 1624 | 88.1 | Q4 |
| | Professional Engineering | Magazine | Trade | English | 1164 | 1167 | 99.7 | Q4 |
| | Journal of Communications Technology and Electronics | Journal | Academic | English | 1113 | 1748 | 63.7 | Q4 |
| | Aerospace America | Journal | Academic | English | 952 | 981 | 97.0 | Q4 |
| | Power | Magazine | Trade | English | 825 | 1062 | 77.7 | Q4 |
| | Fluid Dynamics | Journal | Academic | English | 757 | 937 | 80.8 | Q4 |
| | **Top 10** | | | | **16291** | **19525** | **83.4** | |
| SSCI | Forbes | Magazine | Consumer | English | 3371 | 3428 | 98.3 | Q1 |
| | Fortune | Magazine | Consumer | English | 1470 | 1483 | 99.1 | Q4 |
| | Nation | Magazine | Consumer | English | 1207 | 1571 | 76.8 | Q3 |
| | Actual Problems of Economics | Journal | Academic | Russian | 870 | 2605 | 33.4 | Q4 |
| | New Republic | Journal | Consumer | English | 765 | 863 | 88.6 | Q3 |
| | Sotsiologicheskie Issledovaniya | Journal | Academic | Russian | 593 | 2132 | 27.8 | Q4 |
| | Voprosy Psikhologii | Journal | Academic | Russian | 526 | 868 | 60.6 | Q4 |
| | Internationale Politik | Magazine | Consumer | German | 522 | 961 | 54.3 | Q4 |
| | Commentary | Magazine | Consumer | English | 452 | 693 | 65.2 | Q4 |
| | Library Journal | Magazine | Trade | English | 417 | 788 | 52.9 | Q4 |
| | **Top 10** | | | | **10193** | **15392** | **66.2** | |
| A&HCI | Europe Revue Litteraire Mensuelle | Journal | Academic | French | 2239 | 2310 | 96.9 | n/a |
| | A U Architecture and Urbanism | Magazine | Trade | English | 1821 | 2020 | 90.1 | n/a |
| | Historia | Journal | Academic | French | 1447 | 1604 | 90.2 | n/a |
| | Voprosy Filosofii | Journal | Academic | Russian | 1383 | 1743 | 79.3 | n/a |
| | Ponte | Magazine | Consumer | Italian | 1375 | 1549 | 88.8 | n/a |
| | Connaissance Des Arts | Magazine | Consumer | French | 1202 | 1203 | 99.9 | n/a |
| | Architectural Digest | Magazine | Consumer | English | 1195 | 1198 | 99.7 | n/a |
| | Positif | Magazine | Consumer | French | 1144 | 1144 | 100.0 | n/a |
| | Space | Magazine | Trade | English | 1114 | 1496 | 74.5 | n/a |
| | Architectural Review | Magazine | Trade | English | 1112 | 1223 | 90.9 | n/a |
| | **Top 10** | | | | **14032** | **15490** | **90.6** | |

Notes: Data accessed on 10 June, 2018. Only articles and reviews considered.

[a, b]: We retrieved both serial type and content type from Ulrich's Knowledgebase and matched them based on ISSN.

[c]: For multi-language journals, the language listed is the journal's primary publishing language.

nature of these periodicals.[7] As shown, 11 out of the above 30 periodicals are indeed journals, while 19 are actually magazines. That is to say, the majority of top journals with the problem of missing address information are not scientific

3.6 Origins of completely missing address data

Two possible reasons may lead to the author address data being missing from WoS. One could be that the authors simply did not provide the address data in the original publications, while the other is that the original publications include the address data but for some reason these data are not in WoS. Examining which reason contributes to the missing rate in the address field, as well as to what extent it contributes, is important before proposing possible interventions. Given that Table 3 shows that most journals have a certain number of publications reporting addresses, we then turn to selecting articles rather than journals to examine the reasons for the address information missing from WoS. To better understand the origin of this information omission, we examined both possible causes based on full-text scanning of a sample in the most recent examined year of 2015.

We first downloaded the bibliographic data of all 2015 publications indexed in the WoS Core Collection database (SCIE, SSCI, and A&HCI) which do not have address information. The search query is as follows:

> Query #4: (PY=2015) NOT (AD=(A* OR B* OR C* OR D* OR E* OR F* OR G* OR H* OR I* OR J* OR K* OR L* OR M* OR N* OR O* OR P* OR Q* OR R* OR S* OR T* OR U* OR V* OR W* OR X* OR Y* OR Z*OR 0* OR 1* OR 2* OR 3* OR 4* OR 5* OR 6* OR 7* OR 8* OR 9*))

We confined document types to Article or Review and executed the above search on 11 November 2017; it returned 15,959 hits. We next retrieved the publications' unique IDs, randomly selected 2% of these records (i.e., 320 publications), and downloaded their full texts when possible. We downloaded the full texts of 176 articles after different attempts via WoS, Scopus, and Google. Two coauthors scanned those full texts and coded address information independently based on previously agreed-upon procedures. The agreement rate of inter-coder reliability reached 98.9% on the first round. For those two cases which were in disagreement, the two coauthors had a discussion and reached consensus in the end.

The data show that the two reasons accounting for the address information being missing coexist. Approximately three-fifths of the records (i.e., 105 out of 176 publications) do not have any address information in the full text, while 40% of our sample (71 out of 176) have different types of address information, such as affiliation, city, state, country, or any combination of these reported in the full text. We also found that in our sample there are 15 publications that have complete address information reported, including affiliation, city, state, and country, but for some reason this information is not indexed in WoS.[8] If the sample finding holds for a larger population, it means the problem of completely missing address information can be attenuated with improved indexing on the WoS side alone.

---

[7] www.ulrichsweb.com. Accessed through Fudan University Library on 4 April 2018.

[8] In our sample, one record is actually a book of conference abstracts that is 121 pages long. Though the addresses for presenters are complete, it is understandable that WoS did not index this information (UTID=000361900100001).

3.7 Beyond completely missing data: A case of country name omission

In addition to the problem of completely missing address information, which may lead to making substantial records invisible for analysis, partially missing address data such as missing country name should also be heeded. In this research we selected the U.S., the most prolific country of scientific publications, to illustrate this situation. We chose the U.S. for the following reasons. First, many bibliometric studies utilize the field of country to retrieve and download publication data for research performance evaluations, and thus researchers should be aware of the problem of missing country data. Second, because the U.S. is the most prolific country in many research domains, other countries often use it as the benchmarking case for comparison. Third, based on our trials and errors on the top 20 most productive countries, the U.S. is the only one which has a substantial number of cases with country-name data missing from the address field.

An author's address in WoS typically ends with a standardized country name. However, we found that many publications, including most of the influential research (co)authored by U.S. researchers, do not have country names but have state names in the address field. Take a highly cited paper titled "Electrophoretic Transfer of Proteins from Polyacrylamide Gels to Nitrocellulose Sheets–Procedure and Some Applications" (Towbin, et al., 1979), for example. This paper was published in *Proceedings of the National Academy of Sciences* and coauthored by scholars from Switzerland and the U.S. (see Figure 4). The address field for this paper does not contain the country name of the U.S. That means when researchers search "US or "the United States" or "USA" in the address field they would miss this "heavy hitter" (cited in WoS more than 54 thousand times by 2016).

(Insert Figure 4 here)

This is a single case illustrating partial information omission that may lead to inaccurate and incomprehensive retrieved hits when the search strategy is based on address-relevant fields.

We next investigate how common it is for U.S.-coauthored papers to be without the country name in the address field. Using a composite Boolean search (Query #5), we found that U.S. scholars contributed to 9.26 million SCIE articles and reviews[9] from 1900 to 2015 and that 1.52 million (or about 16%) of them have no country name in the address field.

Query #5: (CU=USA) NOT (AD=USA)

---

[9] The reason we can retrieve the U.S. papers that do not include the country name in the address field (i.e., AD) is because the database provider Clarivate Analytics (formerly Thompson Reuters) created ad hoc a field of country name (CU) and added "USA" if AD contained information of the 50 states of the USA before 1998. When experienced bibliometricians analyze publications by country online via a "CU" search, they will not incur the underestimation problem. Yet it is possible for more novice professionals to search and analyze via the "AD" field. So for them, we need to point out the difference between CU and AD for online analysis. But more important, for researchers who download raw texts of bibliographical data of U.S. publications and plan to do more detailed analysis offline, there is no such field of "CU" in the retrievals. Researchers need to be aware of missing country names and take further steps for cleaning.

Figure 4 Example of a publication without a country name in the address field

Source: WoS. Data accessed on 15 November 2016.

As Panel A of Figure 5 shows, U.S. publications without country names (i.e., CU) exist in the period of 1966–1998, with the missing address rates beginning at 97% in 1966 and dropping to 34% in 1997 then dropping sharply to 0.1% in 1998.

(Insert Figure 5 here)

SSCI is in the same situation as SCIE (Panel B of Figure 5). In the most troubled period of 1966–1997, nearly 0.34 million out of 0.94 million U.S. SSCI publications (articles and reviews) have no country in the address field. The problem almost vanished for social science research after 1997. For the A&HCI database (Panel C of Figure 5), about 31 thousand out of 212 thousand U.S. publications (about 15%) have no "USA" in the address field in the period of 1975–1997, especially in the first four years (from 1975 to 1978).

Figure 5 also shows that the WoS database has improved dramatically since 1998 on the problem of partially missing data. This is good news for bibliometricians who investigate the research landscape involving the U.S. in recent years.

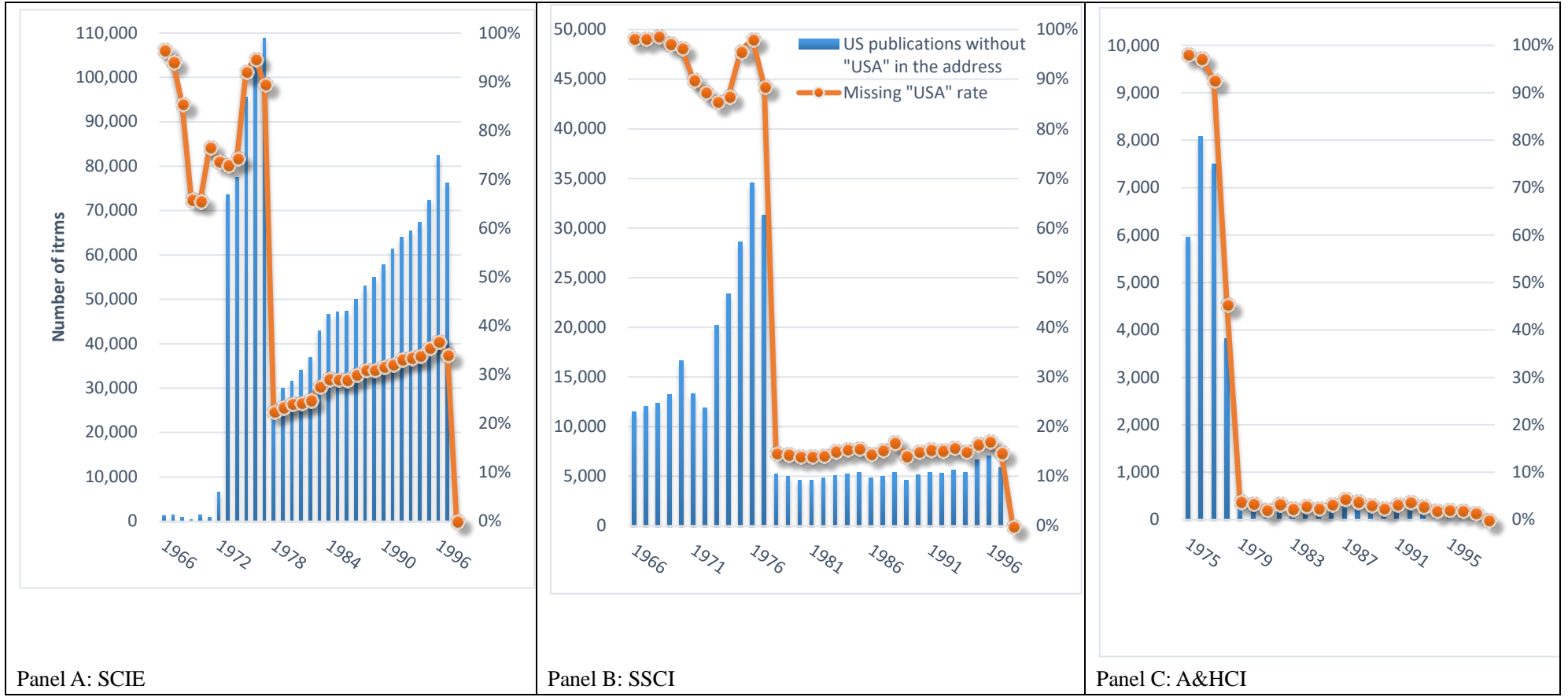

Figure 5 U.S. publications without "USA" in the address field: 1976–1997

Panel A: SCIE    Panel B: SSCI    Panel C: A&HCI

Notes: Data accessed on 22 March 2018.

Only articles and reviews considered.

Blue bars refer to U.S. publications without the country name in the address field. Red dots are their missing address rates.

4. Conclusion and discussion

4.1 Major findings

Utilizing WoS Core Collection database, we probe the status quo and dynamics of completely missing information from the address field between different databases, document types, and publishing languages. We find the problem persists in all three databases over the whole study period (1900–2015). It is particularly serious for SCIE and SSCI indexed publications before the early 1970s, and it remains worrisome for A&HCI publications. Influential articles (i.e., highly cited papers and those published in prestigious journals) also have missing author address information. Based on a random sample of WoS publications from 2015, we further identified two coexisting reasons for the information being completely missing from the address field. We find approximately 60% of articles did not report address-related information in the full text, while 40% do contain some address information that is not indexed by WoS. In addition to completely missing address information, we also examined partially missing address information in the case of the U.S. We find that a large number of U.S. publications during 1966–1997 have no country name reported in the address field. This should be heeded for research evaluation when searching for and retrieving address-relevant data over that period.

4.2 Ignoring the data omission problem

Ignoring the problem of missing data may lead to inaccurate findings or may perplex researchers due to data inconsistency. Take U.S. publications between 1966 and 1996, for instance. The previous section demonstrates that when the search strategy is based on address-relevant fields (CU or AD), being unaware of or ignoring partially missing addresses can lead to an underestimated evaluation of U.S. scientific outputs because searching only in the address field will exclude otherwise viable results from being included simply because they do not include the country name in that field.

Take publishing language for another example. As illustrated in Table 2, when confined to articles and reviews there are more than 35 thousand SSCI publications and 52 thousand A&HCI publications written in English that have no address information, accounting for 2.3% and 16.5% of all SSCI and A&HCI English articles, respectively. In comparison, about one-third of SSCI publications and two-thirds of A&HCI publications written in Russian have completely missing data from the field of author address. That means if a researcher uses the WoS database to investigate the patterns of research written in a specific language, he or she may miscalculate the number, proportions, and rankings of publications of investigated countries, institutions, languages, collaborations, and other geography-related information, given not only the existence of missing data in the address field but also the various missing address rates by different languages.

4.3 Limitations and contributions

This research has two major limitations. First, our examination focuses on the address field of WoS database. Though we conducted some explorations in Scopus, we find that database is not quite ready for this type of analysis at this moment (see the appendix). A comparative study with other databases would be worthy of future work when applicable. Second, though we depict and characterize the phenomenon of missing data in the address field, unfortunately in many cases because the denominator (the total number of publications) is unknown due to the absence of authors' address information (i.e., without knowing what is omitted), we do not know to what extent omission-induced errors impact research conclusions.[10] In this sense, we would like to remind readers to be cognizant of the situation of information omission but also be cautious of claiming its impact in bibliometric assessments.

Bearing these limitations in mind, this research contributes to the ongoing debates on the role of credible data in evidence-based policy evaluation. Within the authors' best knowledge, this is the first publication explicitly stating and differentiating partially and completely missing information from the address field of WoS-indexed publications. Echoing a previous study that states that the WoS database is far from being free from errors (Franceschini, Maisano, & Mastrogiacomo 2016b), our analyses provide evidence that when utilizing WoS data for research evaluation scholars should be cognizant of the problem of missing information and its potential impacts. This is also true for studies on research impact and competitiveness focusing on a specific period, a language, or U.S. research output. We argue that although the problem of missing address information has diminished in recent years, ignoring either completely or partially missing data may lead to inaccurate analyzing samples and thus distortion of bibliometric indicators/metrics.

As noted by Jacsó (2009), the heterogeniety of missing information from the address field is primarily due to the inconsistency of journal policies on address inclusion. This can be traced back to the original rationale of creating a journal publications database: it served mainly for efficient information retrieval and archival purposes rather than for bibliometric-based research evaluation. Our analysis shows the omission of author address information may present a hindrance to effective searching and thus poses potential threats to the credibility of bibliometric results evaluating productivity and impacts of institutions and nations. Search queries using the address field could miss publications without address information. Even though luckily researchers do not use the address field for searching at the data retrieval stage (they typically use the topic field), analysis on address-relevant fields could be worrisome without acknowledging or discussing the impacts of missing data on findings, let alone addressing the problem. This is particularly interesting in data-driven research or evidence-based policy making.

4.4 What we can do when evaluating publications with missing address information?

Given the sheer size of missing address information from WoS as well as the anticipated increasing uses of the address field for research evaluation, we propose the following remedies when encountering these problems.

To begin with, avoid unintentional errors at the information retrieval stage. Our analyses show that information omission demonstrates itself in two types: completely missing data and partially missing data. Neither type of missing data is distributed evenly or randomly by time period, document type, publishing language, or country, and thus induced errors cannot be ignored by assuming they will even out. Researchers can adopt complementary or alternative searching strategies to obtain the relevant data from the very beginning. For instance, search the field of country name (CU) rather than address (AD) for U.S. publications, and combine WoS and other data sources when analyzing and comparing research output prior to the 1970s.

Second, assess the extent of missing data from the retrieved data before conducting any analyses. One technique is to check whether the missing data follow a distribution pattern similar to that of those records with addresses for the key variables being examined. The second option, if possible, is try to supplement missing data with alternative sources.

---

[10] We would like to thank an anonymous reviewer for bringing up this issue.

Last but not least, bear in mind and remind readers of potential impacts of missing information when drawing conclusions and give evidence-based policy suggestions. Data quality is critical for responsible empirical research. Researchers should always check first if the data can speak for themselves. If not, then try remedies and alternative data sources. If these cannot solve the problem, at least the data limitations can be pointed out and the findings can be interpreted with caution.

**Acknowledgment**

This research was funded by the National Natural Science Foundation of China (#71303147) and Zhejiang Provincial Natural Science Foundation of China (#LQ18G030010). We would like to thank Dr. Ludo Waltman and two anonymous referees for their insightful suggestions and comments which have significantly improved the manuscript. The views expressed herein are those of the authors. We are responsible for any errors.

**Appendix**

A robustness check on the problem of missing address data in Scopus

To replicate the analysis we conducted in WoS in another major publication database, we tried two rounds of searches to retrieve Scopus indexed publications without address information. This robustness check reveals two things:

1. The problem of missing address information also exists in Scopus to the extent that it should not be ignored.

2. The current structure of Scopus is not ready for users to "run broad wildcard queries" to retrieve papers with information missing from the address field.

The two rounds of searching queries are as follows.

*Round 1: Search affiliation field*

Query # 6: "PUBYEAR IS 2015"                                           (returned 2,846,251 hits)
Query # 7: "PUBYEAR IS 2015 AND AFFIL (A* OR B* OR C* OR D* OR E* OR F* OR G* OR H* OR I* OR J* OR K* OR L* OR M* OR N* OR O* OR P* OR Q* OR R* OR S* OR T* OR U* OR V* OR W* OR X* OR Y* OR Z* OR 0* OR 1* OR 2* OR 3* OR 4* OR 5* OR 6* OR 7* OR 8* OR 9*)"          (returned 2,686,334 records)

Query 8: "#6 and not #7"                                               (returned 159,917 records)

The above searches were conducted on 19 November 2017. Theoretically, similar to WoS results, the returned 159,917 record hits of Query #7 are those records without address data. Yet a further examination revealed some records in fact included affiliation and country/territory information.

To research this problem, we sent two inquiry emails to the Product Support Engineer of Science Direct & Scopus (Case No. 171120-001076) regarding the missing country/region information on 22 November and 1 December 2017, respectively, and were told that Scopus is "not designed for user to run broad wildcard queries such as the search that you are attempting to perform."

*Round 2: Search country field*

Query # 9: "PUBYEAR IS 2015"                                           (returned 2,845,903 hits)
Query # 10: "PUBYEAR IS 2015 AND (LIMIT-TO (AFFILCOUNTRY, "Undefined "))
                                                                       (returned 196,253 hits)

As suggested by one reviewer, we also tried to solicit Scopus records without country information by limiting the search to the value of "undefined" through the filter of country/territory (please note that the "undefined" option is not available in the field of affiliation). We conducted the searches on 3 April 2018.

As shown this problem of missing address data is non-neglectable in Scopus as well. In 2015 there are 196,253 records which do not have data in the field of country name. However, similar to the round 1 search, a closer examination reveals some records do include affiliation information as demonstrated in the left panel of Figure A1.

Figure A1 Screenshot of publications with "undefined" value in country field

We then limited the affiliation to "Peking University" and clicked the detail pages of the first 20 records, and we found that all of them have complete address information, including affiliation and country name. The finding further echoes the statement of the Scopus Product Support Engineer that the database is not designed for running broad wildcard queries to identify records with information missing from the address field.

Given no reliable way so far to identify and analyze publications missing address information indexed in Scopus online, we decided to focus our analysis on Web of Science in this paper.